\documentstyle[12pt,twoside,fleqn,espcrc1]{article}

\input psfig
\def\3he{$^3$He}
\def\4he{$^4$He}
\def\6li{$^6$Li}
\def\7li{$^7$Li}
\def\he3{$^3$He}
\def\eg{{\it e.g.}}
\def\ie{{\it i.e.}}
\def\etal{{\it et al.~}}
\title{The BBN Manifesto}

\author{Terry P. Walker\address{Departments of Physics and Astronomy, The Ohio State University  \\ 
 174 W. 18th Ave., Columbus, Ohio 43210}%
        \thanks{Invited talk from the Fourth International Conference on Nuclei in the Cosmos, Notre Dame, Indiana, June 20-27, 1996}}

\begin{document}
\maketitle

\begin{abstract}
In this manifesto I 
 review 
the current status of standard big bang nucleosynthesis in light of recent observational data and discuss
 the importance of near-future observations as direct tests of standard 
 BBN.
\end{abstract}

\section{BBN is a Testable Theory}

The status of big bang nucleosynthesis (BBN) as a cornerstone of the hot big 
bang cosmology rests on the agreement between the theoretical predictions and 
the primordial abundances (as inferred from observational data) of the light 
elements deuterium (D), helium-3 (\3he), helium-4 (\4he), and lithium-7 (\7li).
The strength of BBN is that it is a testable theory:  from its beginings in 
 the late 1940's, BBN  predicted a 
primordial \4he abundance of $\sim$25\% by mass\cite{history50}.  Hoyle and Tayler\cite{HT67}
asserted
that stars could be found with no \4he, thereby debunking primordial
nucleosynthesis.  To the contrary, stars  have \4he at the level of 25\% by
mass, thus providing the strongest evidence to date that BBN synthesized 
light elements a few minutes after the big bang.  
The 1980's saw BBN pass the \7li test when the predicted primordial \7li abundance at $10^{-10}$ 
relative to hydrogen was verified via the observation of lithium in metal-poor halo stars 
of our Galaxy\cite{Spites82}.  The good agreement of standard BBN's predictions                             
with the primordial abundances of the light elements as inferred from 
observational data allowed us to accurately bound the baryon density 
of the universe and the expansion rate of the universe at the time of 
nucleosynthesis\cite{YTSSO,WSSOK}.  As such, BBN provided the critical link for the
`astro-particle connection'.   The 1990's have brought a multitude of 
observations related to the primordial abundances of D, \3he, \4he, and \7li.   
In addition to increasing the existing data sets, the recent observations of 
deuterium in three QSO absorption line systems\cite{rh96a,rh96b,tytler96,bt96} show promise of the 
first {\it direct} 
measurement of a primordial abundance and may provide the next critical test 
of
standard BBN.
These improved observations (particularly \4he) seem to indicate that the concordance 
between the predictions of standard BBN, simple models for Galactic and stellar chemical evolution, and the observational data may not be as good as we once
thought\cite{hataii,cstIII}.  However, because the primordial light element abundances are not necessarily 
directly measurable, both the quality of the observational data and of the 
chemical and stellar 
evolution models used to infer the primordial abundances from this data have 
become increasingly important as the observational data sets have grown.  
In most cases the observed abundances of the light elements 
must be corrected for the astrophysical contamination that occurs during 
the evolution of the Universe and therefore {\it the determination of the 
primordial light element abundances is model dependent.}  As a consequence, 
testing BBN theory with inferred primordial observations relies upon models of 
chemical and stellar evolution that trace the fate of the primordial light elements 
through various astrophysical environments.  
With that in mind we discuss below, element by element, the current status 
of the primordial abundances as they are inferred from observational data.  
We discuss the relative importance of the extrapolation of the observational data to the primordial abundances:  as it stands now, most of the uncertainty in determining the primordial abundances of D and \3he lies in complex Galactic chemical evolution models whereas most of the uncertainty in determining the primordial abundances of \4he and \7li lies in the understanding of the regions where the observational data is taken (metal-poor extragalactic HII regions and metal-poor Galactic halo stars, respectively).  In this manifesto we 
 review 
the current status of BBN in light of recent observational data and discuss
 the importance of near-future deuterium observations as direct tests of the standard 
 cosmological model.

\section{Predictions}
\begin{figure}
\psfig{file=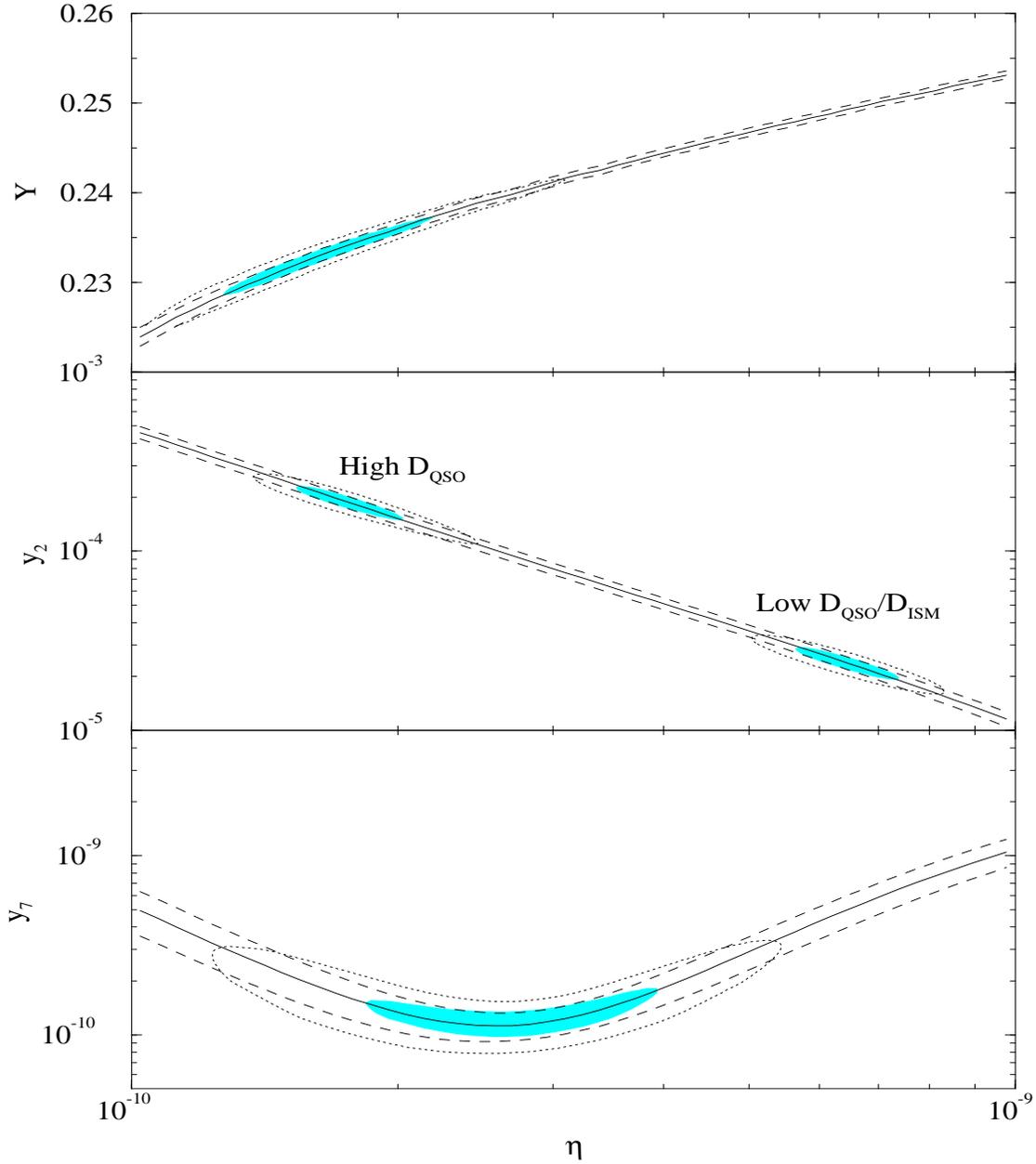,height=7.0in,width=6.0in}
\caption{The standard BBN predictions for the abundances of D,\3he,\4he, and \7li as a function of $\eta$, the baryon-to-photon ratio.  Dashed curves are the $2-\sigma$ uncertainties in the predicted abundances.  Also shown in overlay are the inferred primordial abundances for these light elements and the two possible deuterium abundances from QSO absorption line systems (see text for discussion).}
\label{fig:Fig1}
\end{figure}

The predictions of the standard BBN are uniquely determined by one
  parameter, the density of baryons (parameterized by 
  $\eta$ - the baryon-to-photon ratio), 
provided we assume a homogeneous and isotropic hot big bang and that
  the 
  energy density of the 
  Universe at the time of nucleosythesis (about 1 second after the big bang)
  is described by the standard model of particle physics ($\rho_{\rm tot} =
\rho_\gamma + \rho_e + {\rm N}_\nu \rho_\nu$, where $\rho_\gamma$,
$\rho_e$, and $\rho_\nu$ are the energy density of photons, electrons
and positrons, and massless neutrinos (one species), respectively, and $N_\nu$ is the equivalent number of massless 
  neutrino species which, in the standard BBN is exactly 3). 
  The current predictions of standard BBN are made with essentially the same 
  code as 
developed by Wagoner\cite{Wagoner} in the early 1970's (for a review of the 
current status of BBN predictions, see any of the following:\cite{WSSOK,SKM,CSTI,hatai}).  The
 changes in BBN on the theoretical side have mostly involved refinements and 
 reductions in the uncertainties in the input 
 physics (\eg, nuclear reaction rates and neutron life-time)  
 to the point where, with the possible exception of \7li, the errors in the 
 predictions are much smaller than the uncertainties in the 
 inferred primordial abundances.  In Figure 1 we show the predictions of 
 standard BBN (\ie, assuming a homogeneous and isotropic Universe with 
 3 massless neutrino species) as a function of $\eta$\cite{hataii}.  The width of each 
 curve represents the $2-\sigma$ uncertainty in the predicted abundance.
Increasing $\eta$ corresponds to increasing the nucleon density during 
nucleosynthesis and therefore increasing the efficiency of nuclear reactions, 
both in destroying more D and \3he and in creating more \4he.  \7li falls and rises due 
to a competition between destruction and production reactions.

\section{Observations}

With the predictions of standard BBN well-understood, we turn to the data for 
a critical comparison.  As mentioned above, the primordial abundances are, 
with the possible exception of deuterium (see the discussion of QSO absorption line 
systems, below), obtained from
contaminated data and therefore we must rely on a model for the evolution of a
given element as it is processed thru one or more generations of stars and as
it is enhanced or depleted by galactic processes (\eg, infall of primordial
material or outflow of processed material, respectively).  Our goal is to derive bounds on
the primordial abundances which are as insensitive to the details of their
processing history as possible.  Below find short reviews of the current status
of each of the BBN elements.

\subsection{Helium-4}

Compared to D, \3he, and \7li, the evolution of \4he is the simplest 
to follow.  Stars produce \4he as a
by-product of hydrogen burning and massive
stars produce metals (Carbon, Nitrogen, and Oxygen). Therefore we expect the
amount of \4he in a star and the abundances of CNO nuclei to be
correlated\cite{Piembert}.  Using the \4he abundances from more than 60
metal-poor extragalactic HII regions\cite{Pagel,Skillman,izzy}, 
Olive and Steigman\cite{os96} find from a
least-squares regression:

\begin{equation}
        {\rm Y}_{\rm p} = 0.234 \pm 0.002\, ({\rm stat}) 
                                \pm \Delta {\rm Y}_{{\rm syst}}, 
\end{equation} 
where $\Delta {\rm Y}_{{\rm syst}}$ represents any residual systematic errors in the derived \4he mass fractions.
Other than the correlation of stellar production of \4he  with CNO
production, there is no chemical evolution dependence to this result.  Note that this analysis includes the entire Izotov \etal\cite{izzy} data set of 27 HII regions and has the
same $2-\sigma$ upper-limit to ${\rm Y}_{\rm p}$ as previous estimates\cite{OS94}.  Also note that a simple average the lowest metalicity HII regions yields an identical $2-\sigma$ upper-bound to ${\rm Y}_{\rm p}$\cite{os96}.  The real
issue here is the size of the systematic errors which accompany the conversion
of the intensities of singly ionized \4he emission lines to total mass fractions of
\4he.  In the following discussions, I'll assume  $\Delta {\rm Y}_{{\rm syst}}\sim 0.005$, the level independently estimated by several groups\cite{Pagel,Skillman,OS94,skilldude}, and return to the possibility that it may be required to be larger\cite{CSTI}.

\subsection{Lithium-7}

Next we turn to lithium.  \7li is made in the big-bang at levels greater
than $\sim 10^{-10}$ relative to hydrogen.  Its abundance, as observed in 
about 100 metal-poor ($-3.8\le$ [Fe/H]$\le
-1.3$) halo stars\cite{Thorburn,Molaro95,Spites96} is
roughly constant with respect to metallicity (and for $T_{eff}\ge 5800$K):
\begin{equation}
\left(\frac{Li}{H}\right)_{Halo} = (1.6\pm 0.1)\times 10^{-10}.
\end{equation}
The constancy of this plateau as a function of metallicity and stellar surface temperature is taken as
evidence that the Pop. II halo star lithium abundance is in fact very close to the initial lithium abundance of the gas which formed these stars.  In order to interpret this as the primordial lithium abundance we must ensure that substantial depletion or creation of lithium-7 could not have occured. The halo stars may have started 
with
a higher abundance of lithium and then uniformly depleted lithium down to
the observed plateau abundance\cite{Yale}.  Observations of \6li, Be, and B in these same 
stars
limit the amount of such depletion to be no more than a factor of two\cite{walker93,steigman93} and in
fact 
the models of Vauclair and Charbonnel\cite{cv} that include microscopic diffusion and a
modest stellar wind can account for all of the correlations and dispersions
claimed to exist in the plateau provided the initial abundance of \7li is no
greater than $\sim 3 \times 10^{-10}$ relative to hydrogen.  On the other hand, some of the 
lithium contained in halo stars may have been made by
cosmic ray spallation in the gas prior to their formation.  Again,
observations of Be and B in these stars limits the spallation contribution
to be less than 20\% of the halo abundance\cite{sw92}. 
In addition to the above mentioned systematic uncertainties, there are  systematic uncertainties associated
with the modeling of the stellar atmospheres (estimated at $^{+0.4}_{-0.3}$ dex) And so, we estimate the primordial \7li abundance 
to
be in the range
\begin{equation}
\left(\frac{Li}{H}\right)_p = (1.6\pm 0.1{^{+0.4}_{-0.3}}{ ^{+1.6} _{-0.3}})\times 
10^{-10}. 
\end{equation}

\subsection{Deuterium and Helium-3}

The lower-bound to primordial Deuterium provides the cleanest constraint on BBN.  Since D is destroyed in all 
astrophysical 
environments\cite{els}, any measurement of deuterium places a chemical 
evolution-independent lower bound to the primordial abundance of 
deuterium (and thru the BBN predictions, an upper bound to $\eta$). {\it Whatever
the primordial abundances of \3he, \4he, and \7li, they must be in agreement
with this lower bound to deuterium if standard BBN is to be correct.}  Two 
classes of local deuterium observations exist: the pre-solar nebula abundance 
($\left({\rm D}/{\rm H}\right)_{\odot}$)
 and the abundance in the local interstellar medium (ISM) 
 ($\left({\rm D}/{\rm H}\right)_{ISM}$). The 
 pre-solar 
 abundance of deuterium as inferred from \3he measurements in meteorites 
 and the solar wind\cite{geiss} is\cite{st95}
\begin{equation}
\left({\rm D}/{\rm H}\right)_\odot = (2.6\pm0.9)\times10^{-5}.
\end{equation}
  The ISM 
deuterium abundance is measured with HST as absorption along the line of 
sight to relatively nearby stars\cite{Linsky}:
\begin{equation}
\left({\rm D}/{\rm H}\right)_{ISM} = (1.6\pm 0.2)\times10^{-5}.
\end{equation}

And so, simply using the Universe's inability to synthesize deuterium anywhere
but the Big Bang, we have a model independent lower bound to the primordial
deuterium abundance:
\begin{equation}
\left({\rm D}/{\rm H}\right)_{p} \ge 1.2 \times 10^{-5},
\end{equation}
which corresponds to $\eta\le 8\times10^{-10}$ and is in good agreement with
the inferred primordial abundances of \4he and \7li reported here.

Although the lower-bound to primordial deuterium is the least model dependent, bounding primordial deuterium from above is a different story since 
it necessarily involves a model for 
chemical evolution.  Any gas that is processed thru stars has all of 
its deuterium destroyed.  In addition to providing a deuterium destruction 
factor, models of chemical evolution should also describe observed properties 
of the Galaxy such as the age-metallicity relation, the gas fraction, the 
overall metallicity, individual abundance ratios, the lack of metal-poor G-dwarfs in
the solar neighborhood, and the \3he abundance at 
in the pre-solar nebula and in the ISM , to name but a few.  At present, the 
severity of these constraints as well as various models abilities to fit them 
are in the eye of the beholder.  
Essentially all models that were constructed prior to the need for large deuterium destruction got a factor of $\sim 2$ destruction of deuterium.  Models of Steigman and Tosi\cite{st95} and of 
Fields\cite{fields} are 
consistent with deuterium destruction by a factor of 2-4 while the recent models
of Scully \etal \cite{scov} can deplete deuterium by a factor of ten. The 
difference between the
two approaches can be traced to outflow - large deuterium destruction is 
leveraged agaisnt the over-production of metals because massive stars can
quickly cycle thru gas (thereby destroying deuterium) but only at the expense of
metal production.  This leverage can be decreased by expelling the metals from the Galaxy, as hi-lighted by the stochastic parameterization of Copi \etal\cite{cstII} and explicitly demonstrated in the Scully \etal models\cite{scov}.  It is not clear whether such `outflow models' have anything to do with our galaxy or whether they are better fits to the data than closed models.  Perhaps BBN is showing us the way of the future for chemical evolution models?

The evolution of \he3 is more complicated than that of D.
Not only do we face similar uncertainties shared by D/H stemming from galactic
chemical evolution, but there are considerable uncertainties in the stellar
yields of \he3, particularly in low mass stars.
No one gets the correct answer - \3he is grossly overproduced by the solar
epoch in all standard chemical evolution models\cite{RST-Olive95-Galli95}
and even in cases where the primordial abundances of D and \he3 are set to
zero, an over-production of \he3 is found \cite{dst96}.
In fact, even in the rather extreme models invoking both stellar and galactic
winds, \3he is always overproduced\cite{scotv}.  We are forced to conclude that
the \3he problem cannot simply be solved by chemical evolution alone and that
there must be something wrong with the stellar production and/or destruction of
\he3.  It has been argued 
that an extra mixing mechanism due to diffusion below the 
convective envelope can lead to the destruction of \he3 in low mass
stars while potentially 
 explaining the high $^{13}$C abundance in globular cluster red giants\cite{Charbonel-Hogan-Wasserburg-Weiss-BoothroydS-BoothroydM}.
It may be possible that some set of these non-standard models in which some
or all of the newly synthesized \3he is destroyed in low mass stars can
simultaneously give the correct \3he chemolution (\ie, start with a
primordial \3he which is not too large and still account for the pre-solar and
ISM \3he abundances) {\it and} account for the stellar \3he production as
observed in the PNe {\it and} explain the $^{13}$C anomalies.  In that case the maximum amount of primordial \3he allowed
could give a lower-bound to the baryon density which is consistent with that
derived from D evolution.
Until the problem of \he3 production and/or destruction in low mass stars
is resolved, it is difficult to see how \he3 can be used as a probe of BBN.

There is a way to remove the question of the chemical evolution from the picture
and that is to measure the abundance of deuterium in sufficiently un-evolved extra-Galactic systems. Such abundances have been measured in high-redshift clouds along the line-of-sight to two high-redshift QSOs.  The deuterium in these Lyman-limit absorption line systems appears as blue-shifted (82 km/s) H-like absorption and has the potential of being primordial.  Unfortunately there are currently two different deuterium abundances reported and they are  separated by an order of magnitude.  One group measures along the line-of-sight to QSO 0014+8818\cite{rh96a,rh96b} and reports D/H$\sim 2\times 10^{-4}$.  Another group\cite{tytler96,bt96} has deuterium abundances along two different lines-of-sight (QSO 1937-1009
and QSO 1009+2956) with an abundance roughly 10 times smaller: 
D/H$\sim 2\times 10^{-5}$.  At the time of this talk, there is no way to determine which group, if either, is measuring the primordial deuterium abundance, but something is drastically wrong with either BBN and/or chemical evolution if either of them is.

\section{BBN Scorecard}

So, how does BBN stack up?  The QSO deuterium dichotomy, frustrating in its inability to hone in on the primordial deuterium abundance, proves useful to examine BBN's performance.  Figure 1 shows the predictions of standard BBN with the inferred primordial abundances as discussed above in overlay (the dotted contour corresponds to the $2-\sigma$ range and the shaded region the $1-\sigma$ range).  Also shown are the two QSO deuterium observations. If the low-D QSO abundance is taken to be the primordial deuterium abundance (which is consistent with
vanilla chemical evolution models which predict a factor of 2-4 deuterium depletion), there are problems for BBN if the primordial abundances of \4he and \7li are as reported above.  The predicted mass fraction of \4he consistent with this low-D abundance is $\sim 0.015$ too large and therefore would require a systematic increase in the mass fractions of \4he as extracted from the observations.  This is the `crisis' as identified by Hata \etal\cite{hataii}.  In addition, the \7li abundance consistent with low-D is a factor of 5 greater than the plateau value and would require larger-than-anticipated lithium depletion\cite{Cardell,hataiii}.  Alternatively, the high-D QSO abundance is perfectly acceptable as the primordial deuterium abundance from the point of view of the abundances of \4he and \7li as reported above\cite{fo}.  But, it is a nightmare for chemical evolution.  This is just another manifestation of the `crisis' - namely, the primordial abundance of deuterium required by the inferred primordial abundances of \4he and \7li is a factor of ten larger than that implied by standard chemical evolution arguments starting from the ISM deuterium abundance and working backwards\cite{hataii}.

The resolution of this tension lies with better data and the further removed from chemical evolution, the better.  It is convenient to look towards the QSO absorption line systems but {\it caveat emptor}: it is very difficult to find absorption line systems where nearly primordial D can be meaured.  They must be at high enough redshift to shift the Lyman series into the optical, have high HI column densities, and be `clean' at 82 km/s ({\it i.e.}, the cloud itself must be quiet (low temperature and little bulk motion) and the chance of HI interlopers must be small).  The state of the art in the QSO game will be in
getting high signal-to-noise data so that many lines of the Lyman series can be
used to accurately establish the density of HI.  

\begin{center} Stay tuned ...
\end{center}

\noindent{\bf Acknowledgements}

Thanks to my BBN collaborators whose work I have shamelessly borrowed.  Some of the research presented here was supported by the DOE (DE-AC02-76ER01545).

\end{document}